# Will large language models transform clinical prediction?

Yusuf Yildiz[1*], Goran Nenadic[2], Meghna Jani[3] and David A. Jenkins[1]

## Abstract

**Objective** Large language models (LLMs) are attracting increasing interest in healthcare. This commentary evaluates the potential of LLMs to improve clinical prediction models (CPMs) for diagnostic and prognostic tasks, with a focus on their ability to process longitudinal electronic health record (EHR) data.

**Findings** LLMs show promise in handling multimodal and longitudinal EHR data and can support multi-outcome predictions for diverse health conditions. However, methodological, validation, infrastructural, and regulatory challenges remain. These include inadequate methods for time-to-event modelling, poor calibration of predictions, limited external validation, and bias affecting underrepresented groups. High infrastructure costs and the absence of clear regulatory frameworks further prevent adoption.

**Implications** Further work and interdisciplinary collaboration are needed to support equitable and effective integration into the clinical prediction. Developing temporally aware, fair, and explainable models should be a priority focus for transforming clinical prediction workflow.

## Background

Large language models (LLMs) are rapidly evolving as key technologies in artificial intelligence because of their ability to process, generate, and respond to text. Their popularity rose substantially after the release of OpenAI's generative pre-trained transformer (GPT) [1], and its successful performance across multiple domains [2]. In healthcare, as the capabilities of LLMs begin to be applied to clinical tasks, various applications are now materialising. These include generating discharge summaries, automated clinical coding, answering medical queries, triaging tasks, and synthesising medical literature [3–6].

Recently, the focus on preventive medicine [7], and the potential use of data-hungry algorithms with large-scale electronic health record (EHR) data have increased interest in clinical prediction models (CPMs). CPMs are algorithms that take patient information and provide a probability of a patient experiencing a given outcome currently (diagnostic), or in the future (prognostic) [8]. They rely almost exclusively on using structured and pre-processed EHR data and are often used to improve diagnostic accuracy, predict disease progression, and support clinical decisions [9]. While the ability of LLMs to process multimodal data, understand complex relationships, and achieve high accuracy has raised interest in their potential use as prediction tools, the sensitivity of medical data, along with issues such as bias and fairness in LLMs, may hinder this potential. In this commentary, we discuss the use of LLMs for clinical prediction,

*Correspondence:
Yusuf Yildiz
yusuf.yildiz@postgrad.manchester.ac.uk
[1] Faculty of Biology, Medicine and Health, School of Health Sciences, Division of Informatics, Imaging and Data Sciences, University of Manchester, Manchester, UK
[2] Department of Computer Science, University of Manchester, Manchester, UK
[3] Centre for Epidemiology Versus Arthritis, Centre for Musculoskeletal Research, University of Manchester, Manchester, UK

highlighting the current state of play, the opportunities, and key considerations for the future.

## LLMs in clinical prediction

Healthcare prediction tasks to date primarily use statistical or machine learning methods that apply a given model to the data in a single step. In contrast, LLMs use a two-step development process: pretraining on a large portion of the data to extract general features, followed by fine-tuning, often on a smaller subset for specific outcomes, such as predicting mortality [10] or onset of pancreatic cancer [11]. This approach allows multi-outcome prediction, a promising feature. For example, BEHRT [12], a transformer-based model, predicts the likelihood of 301 conditions in future visits by modelling longitudinal EHR data as chronological sequences of diagnoses and demographics.

Longitudinal nature of EHR data is an important factor to understand patient's history and LLMs show promise to represent this temporality. BEHRT [12] pioneered this by capturing the chronological progression of diagnoses through temporal ordering and including age information of a patient. Additionally, this approach derives the representation from raw or minimally processed EHR [13, 14].

LLMs are versatile and well-suited for handling multimodal data in EHR, including clinical notes, diagnostic codes, lab results, and images [11]. Foresight [15], a GPT-based clinical prediction model, integrates structured and unstructured EHR data to forecast health outcomes across multiple institutions, using virtual patient representations for longitudinal forecasting and demonstrating transportability through external validation.

LLMs can also be trained with different medical terminologies (e.g. International Classification of Diseases (ICD), Systematized Nomenclature of Medicine Clinical Terms (SNOMED-CT), and Read Codes) and have the potential to learn and utilise the hierarchical structure of EHR data regardless of the terminology, potentially improving predictive performance. Med-BERT [11] is trained using both ICD-9 and ICD-10 codes together and validated on the prediction of prolonged length of stay. However, the added value of the hierarchical information and how best to provide the model with this information to learn from it has yet to be explored.

Recent studies further demonstrate LLMs' potential in clinical prediction. CPLLM [16] and NYUTron [17], language models pretrained on clinical notes and EHR data, have demonstrated strong performance in external validations across outcomes such as hospital readmission, diagnostic classification, and mortality prediction. Similarly, MOTOR [18], a foundation model designed for time-to-event analysis, demonstrates strong performance across 19 tasks on three databases. Notably, MOTOR's performance improved when fine-tuned on hospital-specific data, showing the ability for improved transportability. However, findings also show locally trained ML models are more effective than non-fine-tuned LLMs, suggesting the importance of LLMs to be fine-tuned on the setting that they are deployed [19].

The use of LLMs in healthcare is a growing field that has shown significant promise and LLMs have begun to be used within healthcare settings [6]. However, challenges and issues are being ignored at present. These models are still nascent for clinical prediction, and their potential impact on decision making and clinical utility still needs investigation. In the next sections, we discuss the limitations and current challenges.

## Challenges for LLMs in clinical prediction

LLMs show promise for improving clinical prediction by using EHR data to make better diagnostic and prognostic decision-making. However, we have highlighted key challenges that include handling time-to-event data, robust validation, managing high costs, and managing regulations that need to be addressed to enable safe, fair, and effective integration.

### *Methodological gaps*

The adoption of LLMs, which differ in training approaches from standard machine learning applications, presents unique methodological challenges in clinical prediction. Many CPMs have been developed using time-to-event models [20]. While LLMs can process temporal information from longitudinal EHR data, effectively incorporating exact event timing remains a challenge. Current approaches, such as using time tokens or age embeddings, struggle to address survival analysis needs, including censoring (patients lost before an event) and competing events (alternative outcomes preventing the primary event) [21]. Temporal drift, where EHR patterns change over time due to evolving clinical practices, further complicates longitudinal modelling. To address these, researchers can integrate deep learning architectures with survival analysis techniques [22], and more work can be done with time-aware transformer models to model time-to-event outcomes while accounting for censoring [18].

As mentioned earlier, due to the unique two-step development process of LLMs, ensuring a proper data split is crucial to prevent overfitting. Using the same data for both pretraining and fine-tuning can significantly impact the model's precision. Different data-splitting methods have been applied across various LLMs, but the effects of these splits remain underexplored. Figure 1 illustrates different data-splitting approaches.

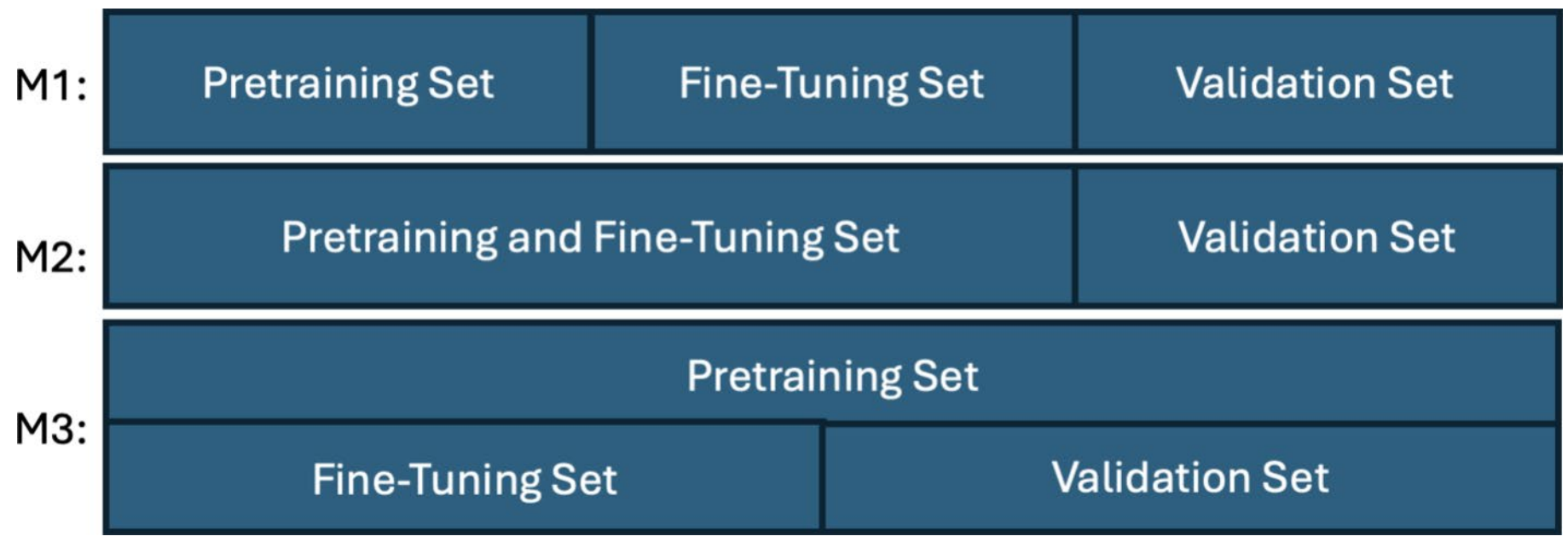

**Fig. 1** Different data-splitting methods in the current literature. M1 Separate portions of the data are used for pretraining, fine-tuning and validation. M2 The same data are used for both pretraining and fine-tuning. M3 The entire dataset is used for pretraining and then fine-tuning, and validation is performed

Another key methodological issue is the concern over unbiased and fair model development. LLMs rely heavily on their training data, and if certain populations are underrepresented or specific treatments are overemphasised, the resulting predictions may be biased. These models require a minimum amount of patient data to accurately identify patterns and representations, but it is unclear if this requirement itself introduces bias. For example, individuals from deprived areas and certain ethnic minorities may interact with healthcare less frequently, potentially leading to poorer model performance for these groups and, consequently, the risk of widening health inequalities. This is especially concerning in healthcare, where the "black box" nature of LLMs adds further complexity. While established methods exist to assess fairness in predictions, these approaches rely heavily on thorough model validation [23]. LLMs must be trained on diverse datasets, and the evaluation phase should include a thorough assessment of bias and fairness in predictions, as discussed next. Table 1 summarises the methodological challenges in applying LLMs to clinical prediction, highlighting current practices, gaps, and proposed solutions for robust model development.

### *Validation and explainability*

Many LLMs are trained for next disease prediction, and these models present unique challenges due to their complexity and multi-outcome ability. In contrast, other methods are often trained to predict the occurrence of an outcome within a specified timeframe. This raises unresolved questions about how to compare these models effectively. Additionally, it is unclear how clinically useful next-disease prediction is, as it is rarely the primary focus of clinical queries or interventions. Significant work is needed to develop complex, clinically relevant time-based prediction tasks. Beyond accuracy, new validation methods are essential to assess the usefulness of multi-outcome predictions in clinical settings. This emphasises the need for improved methods of time representation within these models.

Cross-validation and bootstrapping methods are commonly used for internal validation in statistical models to prevent overfitting and ensure representation. However, these methods have not yet been applied to LLMs, possibly because of their two-step process or potential computational cost. Currently, LLMs rely on a subset of data for development, which is suboptimal.

Calibration, how well a model's predicted probabilities reflect actual outcomes, remains a critical gap and is important for clinicians' trust [24]. As LLMs often show high confidence in incorrect answers, this can lead to unsafe or misleading clinical decisions. Recent studies have shown that many LLMs used in biomedical contexts are substantially mis-calibrated [25]. To ensure safe deployment, validation of LLMs in healthcare should routinely assess calibration and incorporate correction techniques to align model confidence with real-world reliability [25, 26].

**Table 1** Summary of methodological challenges and recommendations for LLMs in clinical prediction

| Current practice | Current gaps | Potential solution |
| --- | --- | --- |
| Use of age embeddings, sequential event ordering, and time tokens | No handling of censoring or competing events; Limited suitability for survival analysis; over/under-representation of time | Develop time embedding techniques for survival analysis; hybrid models with survival analysis (e.g. Cox regression + LLM) |
| Minimum visit count for inclusion | Bias toward high utilisers; underrepresentation of disadvantaged groups | Usage of diverse datasets; bias and fairness evaluation frameworks |

A major barrier to validating LLMs in clinical settings is their opaque decision-making processes. Without clear explainability mechanisms, it is difficult to assess how LLMs reach their predictions, complicating validation and limiting trust among stakeholders. Beyond accuracy metrics, efforts should focus on understanding when and why models produce incorrect predictions, ideally with concrete examples. To build trust and speed up clinical integration, LLMs must incorporate explainability mechanisms that allow transparent evaluation and reduce potential biases, ultimately supporting fair predictions. Establishing transparent evaluation frameworks will improve the adoption and effectiveness of LLMs in clinical settings [27]. Table 2 outlines the challenges in validating LLMs for clinical prediction, including gaps in fairness evaluation and potential solutions for enhancing explainability and clinical utility.

#### *Infrastructure gaps*

The implementation of LLMs in healthcare is expensive, and requires significant computational resources, which can limit their widespread adoption, especially in low-resource settings which can contribute to the healthcare inequality [28]. Cloud-based solutions offer a potential way to overcome these infrastructure challenges, providing healthcare providers with scalable AI capabilities with less need for local infrastructure; however, this can also raise concerns about data privacy and security [29]. Federated learning, an emerging approach where models are trained across decentralised datasets without sharing sensitive data, could address privacy concerns and reduce computational costs [30]. This approach has shown success as a clinical decision support model in COVID-19 patients [31] and initial studies show promise usage of federated learning in transformer models for clinical prediction [32].

Strategic collaborations between healthcare institutions, private companies, and government agencies could help distribute the costs and expertise required for LLM adoption. Public–private partnerships could provide access to cloud infrastructure, high-performance computing, and cybersecurity expertise, making these technologies more accessible and scalable across diverse healthcare environments [33]. Given the reliance on cloud-based services, current GDPR compliance protocols may also need updating to align with LLM technology requirements. Table 3 presents the infrastructure challenges for LLM adoption in healthcare, along with gaps and strategies to promote equitable access and data privacy.

#### *Regulation and trust*

The rapid pace of development for these models has far outstripped the speed at which laws and regulations governing them have been established. As a result, there are currently no well-defined legal frameworks in place to evaluate these models effectively [34, 35]. Prediction models are still often classified as medical devices by drug regulators because they generate outputs that can drive a clinical decision [36]. Current review and authorisation criteria for prediction models are often not detailed [37] and experts believe that the bar is too low especially for bias evaluation and fairness [36].

**Table 2** Challenges in model validation and explainability for LLM-based clinical prediction

| Current practice | Current gaps | Potential solution |
| --- | --- | --- |
| Binary prediction tasks; varying validation metrics | Lack of bias/fairness evaluation; limited clinical usefulness for multi-outcome predictions | Develop standardised metrics for multi-outcome validation; apply fairness tests across subgroups (e.g. race, age) |
| Subset-based validation | Suboptimal data splits; no cross-validation/bootstrapping | Use techniques approximating cross-validation |
| Black-box decision-making | Lack of transparency, no explainability | Develop explainable AI methods; track reasons for prediction failure, use visualisation techniques for embeddings |

**Table 3** Infrastructure-related challenges and mitigation strategies

| Current practice | Current gaps | Potential solution |
| --- | --- | --- |
| Resource-intensive training; reliance on expensive hardware | High costs limit adoption in low-resource settings; risk of healthcare inequality | Foster public–private collaborations; adopt other technologies (i.e. federated learning) |
| Limited digital infrastructure in some hospitals | Data privacy and GDPR compliance concerns | Implement secure cloud systems; update GDPR protocols for LLM-specific needs |

**Table 4** Regulatory and ethical considerations in LLM deployment for clinical prediction

| Current practice | Current gaps | Potential solutions |
|---|---|---|
| Prediction models classified as medical devices. Unclear legal liability for model errors | Lack of LLM-specific regulations, accountability is not defined | Develop tailored regulatory frameworks. Define responsibility clearly (clinician, developer, other stakeholders), involve stakeholders in co-design |
| Limited adherence to reporting guidelines | Lack of transparency | Adopt TRIPOD-LLM guidelines for reporting the studies; create educational programs on LLM ethics |

Additionally, LLMs have different characteristics than current prediction models do, which require a more tailored approach for regulatory oversight. A comprehensive, adaptive, and holistic regulatory framework is needed to keep pace with the evolving capabilities of LLMs and ensure their responsible and ethical use across various domains. Recognising this need to adapt to a developing landscape, educational programs need to be developed that focus on LLM healthcare regulations, ethics, and law.

A key aspect of improving regulatory oversight is improving transparency and reporting. There is a clear lack of reporting since almost none of the studies that used the LLM architecture for clinical prediction have cited any reporting guidelines. A lack of appropriate reporting increases concerns about transparency and thoroughness and reduces the degree of trust and reproducibility of the studies.

The last critical regulatory challenge is determining liability when LLMs make errors. Should the responsibility fall on the healthcare provider, the software developer, or the data provider? Establishing clear accountability is essential for LLM adoption, and regulatory frameworks must address liability concerns to protect both patients and healthcare professionals. Involving all stakeholders, including health professionals, in the development of LLMs will be key to ensuring their responsible use and minimising legal risks. Table 4 summarises the regulatory and trust-related challenges for LLMs in clinical prediction, identifying gaps and recommending solutions for ethical adoption.

## Conclusion

LLMs show promise for improving clinical prediction by using EHR data to make better diagnostic and prognostic decision-making. However, challenges like handling time-to-event data, validating models properly, managing high costs, and sorting out regulations need to be tackled to enable safe, fair, and effective integration.

Future development should prioritise building hybrid models that combine LLMs with traditional statistical approaches, such as Cox regression [22], to better account for censoring, patient dropout, and competing risks. The fine-tuning step should involve training on diverse, representative datasets to reduce biases that may affect underserved populations, including ethnic minorities and individuals from socioeconomically disadvantaged areas. Infrastructure-efficient strategies like federated learning can also reduce costs and improve data privacy.

Explainability and clinical usefulness should be integral to both methodological and validation frameworks. Adding clear explanations, like visualisations of how predictions are made, or how wrong predictions are generated will also help clinicians trust and use these models [27].

Standardised reporting using emerging guidelines such as TRIPOD-LLM [38] will be crucial for improving transparency, reproducibility, and methodological comparability across studies. Engaging interdisciplinary researchers and patient-public involvement can help produce more equitable and acceptable solutions. Providing clear documentation tailored to different user groups will further support this implementation.

As the potential of LLMs continues to be explored, maintaining a strong focus on patient-centred outcomes will aid in smoother integration. While LLMs are not yet transforming clinical prediction, the pace of progress suggests that reality may soon arrive.

### Abbreviations

| | |
|---|---|
| BEHRT | Bidirectional encoder representations from transformers for electronic health records |
| CPMs | Clinical prediction models |
| EHR | Electronic health record |
| GPT | Generative pre-trained transformers |
| ICD | International Classification of Diseases |
| LLMs | Large language models |
| SNOMED-CT | Systematized Nomenclature of Medicine Clinical Terms |

### Acknowledgements
Not applicable

### Authors' contributions
DAJ, MJ and YY conceived the concept of the paper. YY drafted and wrote the main manuscript. GN, DAJ, and MJ provided critical feedback and revisions. All the authors read and approved the final manuscript.

**Funding**
The authors declare that financial support was received for the research, authorship, and/or publication of this article. Yusuf Yildiz was funded by the Republic of Türkiye Ministry of National Education. Meghna Jani is funded by a National Institute for Health and Care Research (NIHR) Advanced Fellowship [NIHR301413]. The views expressed in this publication are those of the authors and not necessarily those of the NIHR, NHS, or the UK Department of Health and Social Care.

**Data availability**
No datasets were generated or analysed during the current study.

## Declarations

**Ethics approval and consent to participate**
Not applicable.

**Consent for publication**
Not applicable.

**Competing interests**
The authors declare no competing interests.

## Publisher's Note